\documentstyle[12pt]{article}

\def\la{\lambda}

\def\ch{\chi}
\def\ps{\psi}

\def\Ga{\Gamma}

\def\cl{{\cal L}}
\def\fr#1#2{{{#1} \over {#2}}}
\def\prt{\partial}

\def\vev#1{\langle {#1}\rangle}

\def\frac#1#2{{\textstyle{{#1}\over {#2}}}}

\def\lsim{\mathrel{\rlap{\lower4pt\hbox{\hskip1pt$\sim$}}
    \raise1pt\hbox{$<$}}}
\def\gsim{\mathrel{\rlap{\lower4pt\hbox{\hskip1pt$\sim$}}
    \raise1pt\hbox{$>$}}}
\def\sqr#1#2{{\vcenter{\vbox{\hrule height.#2pt
         \hbox{\vrule width.#2pt height#1pt \kern#1pt
         \vrule width.#2pt}
         \hrule height.#2pt}}}}

\newcommand{\beq}{\begin{equation}}
\newcommand{\eeq}{\end{equation}}
\newcommand{\bea}{\begin{eqnarray}}
\newcommand{\eea}{\end{eqnarray}}
\newcommand{\rf}[1]{(\ref{#1})}

\begin{document}
\titlepage

\begin{flushright}
{DF/IST-3.97\\}
{March 1997\\}
\end{flushright}
\vglue 2cm
	
\begin{center}
{{\bf Lorentz Invariance and the Cosmological Constant\footnote{This essay 
was selected for an Honorable Mention by the Gravity Research Foundation, 
1997} \\}
\vglue 1.5cm
{O.\ Bertolami\\}
\bigskip
{\it Instituto Superior T\'ecnico,
Departamento de F\'\i sica,\\}
\medskip
{\it Av.\ Rovisco Pais 1, 1096 Lisboa Codex, Portugal\\}
}
\vglue 2cm

\end{center}

\centerline{\bf  Abstract}
\vglue 1cm
\noindent
Non-trivial solutions
in string field theory may lead to the spontaneous breaking of
Lorentz invariance and to new tensor-matter interactions.
It is argued that
requiring the contribution of the vacuum expectation values of  
Lorentz tensors to account for the vacuum energy up to the level that
$\Omega_{0}^{\Lambda} = 0.5$ implies the new interactions range is
$\lambda \sim 10^{-4} \ m$. These conjectured violations of the  
Lorentz
symmetry are consistent with the most stringent experimental limits.

\vfill
\newpage

\setcounter{equation}{0}
\setcounter{page}{2}

\baselineskip=20pt

It is a well known fact that there is no version of the Higgs  
mechanism
in gravity. Indeed, the breaking of Lorentz invariance via  
non-vanishing
vacuum expectation values of Lorentz tensors does not give origin to  
mass
terms to the graviton as this field couples only through derivative
interactions. This feature creates a difficulty in
multidimensional gravity
theories as the extra dimensions components of the graviton will, in
4 dimensions, give origin to long-range interactions, which are  
severely
constrained by observation.
This problem may be circumvented for instance, assuming the
compactification process respects certain requirements, such as for  
instance,
the condition the dimensional reduction procedure involves compact  
manifolds that are coset spaces and accordingly $K-$symmetric metric  
and matter fields
\cite{manton,dp} (see \cite{kmrv,kz} for thorough discussions and a
complete set of references). However, this of course
cannot be regarded as a dynamical explanation. In the context of
string theories, in particular, the spontaneous breaking of the
Lorentz symmetry can be envisaged as an interesting way out  
\cite{ks1,ks2} as string field theory contains interactions that are  
not present in
Kaluza-Klein type theories and their generalizations.
As pointed out in Refs. \cite{ks1,ks2}, this breaking is rather  
natural
in some string theories, such as in the bosonic string where the  
tachyon is present. This breaking may also occur
in the superstring if the dilaton acquires a non-vanishing  
expectation value and its
interactions give origin to negative quadratic couplings for tensors,
possibly non-perturbatively. Notice however, that arguments
based in the effective  
dilation potential suggesting the need of strong coupling \cite{ds},  
and of the implied presence of those interactions in the superstring,   
do not hold once $S-$duality symmetry is
imposed in the effective theory through demanding modular
invariance of the $N=1$ supergravity superpotential \cite{filq}.

It is worth mentioning that the idea of dropping the Lorentz  
invariance
has been previously considered. Indeed, a background or constant
cosmological vector field throughout space has been suggested as a  
way to
introduce into the physical description
our velocity with respect to a preferred reference frame
\cite{phillips}. It has also been
argued, based on the behaviour of the renormalization group
$\beta-$function of non-abelian gauge theories,
that Lorentz invariance may be actually just a low-energy phenomenon
\cite{nn}.
Finally, closer to our discussion, theories of gravity in more than
4 dimensions that are not locally Lorentz invariant have been  
considered
in order to obtain light fermions in chiral representations
\cite{weinberg1} as an alternative to introducing extra gauge fields  
in topologically
non-trivial vacuum configurations \cite{witten1}.

The  breaking of the Lorentz symmetry due to
non-trivial solutions of string field theory is of course related  
with the breaking of the CPT symmetry \cite{kp1}. Interestingly, this  
possibility can, in principle, be verified experimentally in the  
$K^{0}-\bar{K}^{0}$ \cite{kp2}, B and D \cite{ckv}
systems\footnote{The CPT violating effects discussed here are  
unrelated with the ones due to the presence of non-linearities in  
quantum mechanics, presumably induced by quantum gravity, that were  
recently searched by the CPLEAR Collaboration \cite{adler}}. Moreover, this  
violation of the CPT symmetry also allows for an explanation of the  
baryon
asymmetry of the Universe as the tensor-fermion-fermion interactions  
(see below) give
origin, after the breaking of the Lorentz and
CPT symmetries, to a chemical potential that creates in equilibrium a
baryon-antibaryon
asymmetry in the presence of baryon number violating interactions
\cite{bckp}.

The main argument in favour of the breaking of the Lorentz invariance  
arises
in the bosonic open string field theory where interactions
are trilinear type \cite{witten2}. The static potential in this
string field theory proposal has the following form \cite{ks1}:
\begin{eqnarray}
V(S^{i}, T^{i}_{M}) & = & \frac{1}{2} \sum_{i,j} m^{2}_{ij} S^{i}  
S^{j} +
\frac{1}{3!} \sum_{i,j,k} g^{SSS}_{ijk} S^{i} S^{j} S^{k} +
\frac{1}{2} \sum_{i,j} M^{2}_{ij} T^{i}_{M} T^{jM} \nonumber \\
& + &\frac{1}{2} \sum_{i,j,k} g^{STT}_{ijk} S^{i} T^{j}_{M} T^{kM} +
\frac{1}{3!} \sum_{i,j,k} g^{TTT}_{ijk} T^{i}_{MN} T^{jM} T^{kN}
\quad ,
\label{trilinear}
\end{eqnarray}
where $S$ and $T$ represent generic scalar and tensor fields, the  
indices
$M, N$ denote the set of Lorentz tensor indices, $m^{2}_{ij}$ and  
$M^{2}_{ij}$
are the scalar and tensor mass-squared matrices and $g^{SSS}_{ijk}$,
$g^{STT}_{ijk}$ and $g^{TTT}_{ijk}$ the relevant coupling constants.

As discussed in Ref. \cite{ks1}, after expanding the string field in  
terms of
the tachyon  $\phi$, vector $A_{\mu}$ and tensor fields in the
Siegel-Feynman gauge, one obtains the following static potencial:
\beq
V(\phi,A_{\mu}, ...) =  - {\phi^2 \over 2 \alpha'} + a g \phi^3 + b g  
\phi
A_{\mu} A^{\mu} + ...
\quad ,
\label{tachvector}
\eeq
$a$ and $b$ being order one constants and $g$ the on-shell  
three-tachyon
coupling. The vacuum is then clearly unstable and
neglecting loop contributions and other scalars, this instability
gives origin to a
mass-square term to $A_{\mu}$ which is proportional to $\vev{\phi}$.  
If
$\vev{\phi}$ is negative, then the Lorentz symmetry itself
is spontaneously broken.
The vacuum expectation values of any scalars
in the static potential are all of order $M \equiv (\alpha' g)^{-1}$,
a mass scale which is presumably close to the Planck mass, and
the resulting contribution to the vacuum energy
is of order $M^4$ \cite{ks1}. Hence, string interactions involving  
the
tachyon give origin to large violations of the Lorentz invariance,
which are not observed (see below)
as well as to large contributions to the cosmological constant, a
fact that is bluntly contradicted by large scale observations.

Supposedly, the solution for the above embarrassment must be found in  
the field theory of closed strings, where the graviton is present.  
Unfortunately,
little is known about that theory and its relation to the yet even
more fundamental M-Theory. Nevertheless, as far as the Lorentz  
symmetry is concerned, it seems reasonable to assume that  
scalar-tensor-tensor interactions will also be present in closed  
string field theory and, if these have the appropriate sign,  
Lorentz-symmetry breaking ensues in this theory as well. In any case,  
no mechanism is known on how to explain the way contributions to the  
vacuum energy from large vacuum expectation values are cancelled such  
that compatibility with observations are achieved (for a discussion  
of the most salient (failed) attempts see \cite{weinberg2} and  
\cite{weinberg3} for an  update). String theory has however, some  
features as well as symmetries that put this difficulty in a somewhat  
different standing. Strings are non-local objects and that gives  
origin, as depicted in \rf{trilinear}, to interactions that are not  
present in renormalizable field theories. It is conceivable that  
non-localities and duality symmetries present in string theories may  
provide the underlying ``awareness'' \cite{coleman, bb} the high-energy  
theory seems to have in order to account for the vanishing or almost  
perfect cancelling of contributions to the cosmological term at low  
energies. In this respect, it has been recently suggested by Witten  
\cite{witten3} the possibility of relating, via a duality  
transformation, a 4-dimensional field theory to a 3-dimensional one  
with the advantage that in the latter the breaking of supersymmetry,  
a condition imposed upon by phenomenology, does not imply that  
$\Lambda \not= 0$. Furthermore, at least at one-loop in string  
perturbation theory, the so-called Atkin-Lehner symmetry ensures the  
vanishing of the cosmological constant up to that level  
\cite{moore}. These features seem however, insuficient to guarantee,  
to our present knowledge of string theory, the complete vanishing of  
contributions to the potential at the vacuum, and for all purposes  
what is required is a quite radical mechanism to reduce the vacuum  
energy by about 120 orders of magnitude.

Another distinct feature of string theory is the presence of scalar  
fields, the moduli, which like fields lying in the hidden sector of  
supergravity theories interact only gravitationally and, in case of  
being light, may live long enough to dominate the energy density of  
the Universe. This is the well-known Polonyi problem which is closely  
related with the issues of supersymmetry breaking and inflation  
[22, 25-33]. In what follows we shall assume that the mechanism  
accounting for the vanishing or nearly vanishing of the vacuum energy   
does cancel the contributions of all scalar fields to the vacuum  
energy.

Nevertheless, if from the theoretical point of view on one hand it  
does seem that a
symmetry is missing in order to explain the cancellation of the  
cosmological constant  by many orders of magnitude at quite different  
scales, from the observational side on the other, there is mounting  
evidence of a residual non-vanishing vacuum energy, that may be even  
substantial, although not dominating, when compared to the  
contribution of matter in the Universe. Indeed, a non-vanishing  
cosmological constant contributing to  the density parameter, by  
$\Omega_0^{\Lambda} = 0.8$ (together with $\Omega_0^{CDM} = 0.15$ and
$\Omega_0^{Baryons} = 0.05$) allows for the power spectrum of the  
Cold
Dark Matter (CDM) model
to be consistent with both COBE and IRAS observations \cite{esm}  
and also
to make the age of the Universe compatible with the age of the oldest  
globular clusters $t_0 = (15.8 \pm 2.1)~Gyr$ \cite{bh} for $0.58  
< h_{0} < 0.76$~\footnote{Values in this range are consistent with the  
ones arising from studies using the Type I supernovae \cite{rpk} as standard  
candles that indicate $h_0 = 0.67 \pm 0.07$, but are somewhat lower than
the values emerging from studies using classical Cepheid variables  
observed by the Hubble Space Telescope, $h_{0} = 0.82 \pm 0.17$  
\cite{freedman}.}. Furthermore, a non-vanishing upper bound for the  
cosmological constant also arises from
gravitational lensing studies, $\Omega_0^{\Lambda} < 0.75$
\cite{ft,mr}, and is consistent with the Cosmic Background  
Radiation (CBR) power spectrum for $\Omega_0^{\Lambda} = 0.65$  
\cite{os} as well as with recent results of the Supernova  
Cosmology Project suggesting that  $\Omega_0^{\Lambda} < 0.51$  
\cite{perlmutter}. Actually, strategies to determine $\Lambda$ from the  
anisotropies of the
CBR, thanks to the Rees-Sciama effect, have also been proposed  
\cite{ct}. Finally, it has been argued that observations from  
various quarters, such as for instance, from nucleosynthesis and  
X-ray studies of the amount of baryons in rich clusters of galaxies  
to the age of Universe crisis, suggest that $\Omega_0^{\Lambda} = 0.6  
- 0.7$ (and  $\Omega_0^{Matter} = 0.3 - 0.4$ with $h_{0} = 0.7 - 0.8$)  
\cite{kt}.

In order to relate the theoretical possibility of spontaneous  
breaking
of the Lorentz symmetry with the observational contraints and also  
with the
nearly complete cancellation of the vacuum energy we parametrize the  
vacuum
expectation values of the Lorentz tensors as suggested in Ref.  
\cite{kp2}
when studying the limits on the violation of the CPT symmetry:
\beq
\vev{T} = c\left({m_l \over M} \right)^l M
\quad ,
\label{lorentztens}
\eeq
where $c$ is an order one constant, $l$ is a non-negative integer and  
$m_{l}$ a light energy scale when
compared to $M$.

We now assume that the yet unknown mechanism to solve the  
cosmological
constant problem cancels out all contributions from scalars to the  
vacuum energy
of order $M^4$, (and actually contributions of order $M_{GUT}^4$,
$M_{SUSY}^4$, $G_{F}^{-2}$, etc ...) while leaving uncancelled  
contributions of order $ m_{V}^4$, possibly through some soft  
symmetry breaking effect, where we assume consistently with the  
previous discussion that $\Omega_0^{\Lambda} = 0.50$ or in terms of  
energy density $\rho_{V} = (2.39 \times 10^{-3} \ \sqrt{h_0} \ eV)^4  
\equiv m_{V}^4$.  Hence the surviving contributions to the vacuum  
energy from \rf{trilinear} are: $m_{T}^{2} M^2 (m_{l}/M)^{2l}$, $M^4  
(m_{l}/M)^{2l}$ and $M^4 (m_{l}/M)^{3l}$, where $m_T^2$ is the  
mass-squared (a positive eigenvalue of $M^2_{ij}$ in \rf{trilinear} after  
the Lorentz symmetry breaking) of a tensor field arising from the  
extra dimensional components of $T$, the coupling constants  
$g^{STT}_{ijk}$ and $g^{TTT}_{ijk}$ are all order  
one in $M$ units and  $<S> ~\sim M$. A quite  interesting possibility  
then follows supposing the relevant light scale
is the one given by the mass, $m_T$, of the tensor field and that the  
contribution to the vacuum energy due to the non-vanishing  
expectation values of Lorentz tensors saturates $m_{V}^4$, that is:
\beq
m_{l} \equiv m_{T} = m_{V}\quad ,
\label{masstens}
\eeq
for $l = 1$ when taking the third term in eq. (1) or $l = 2$ 
when considering the last two terms in (1). As we shall see below, 
the case $l=1$ is  
compatible with  experiments designed to detect deviations from the  
Lorentz invariance.
This implies that the contributions from the vacuum expectation
of Lorentz tensors can account for the observed vacuum energy if
$m_{T} \sim m_{V}$ for $l = 1$. In this context, the cosmological  
constant puzzle
consists in explaning the conditions under which the cancellation of  
zeroth order terms and of scalar fields do occur. The preceeding  
assumptions lead to a fairly interesting possibility
from the phenomenological point of view as from these new  
intermediate range interactions with $\lambda = m_{T}^{-1} = 8.26 \  
h_{0}^{-1/2} \times 10^{-5} \ m$ are expected. This range is
essentially untested, for pratically any coupling strength, by  
experiments designed
to test putative new finite range interactions \cite{fit} and is  
not, so far, ruled out.

Before we discuss the strength of the tensor-matter interactions
arising from the
breaking of the Lorentz symmetry and the related issue of quantifying
the resulting violating effects in order to confront with  
observation, we mention that the
possibility of saturating the vacuum energy with the contribution of  
a (scalar)
field associated to some new symmetry has been previously
discussed in Ref. \cite{beane}. It was also pointed out in that  
reference that cryogenic mechanical oscillator techniques \cite{price}  
may allow for an improvement of existing limits on Yukawa type
interactions by a factor up to $10^{10}$ in the range about $100 \  
\mu m$.
This range seems also to be favoured for scalar field interactions in  
certain
classes of supersymmetric theories \cite{dg}.

Of course, parametrization (3) gives already an estimate of the  
Lorenz symmetry violating effects, through the ratio $<T>/M$, which  
can be compared with experimental limits. However, in order to better  
quantify this we have actually to consider the
coupling of Lorentz tensor fields to fermions via the trilinear  
string field
interactions. These are actually, the relevant couplings when  
studying CPT
violation \cite{kp2,ckv} and baryogenesis \cite{bckp}. The most  
general form for these interactions is the following:
\beq
\cl_I = \la \fr{T}{M^k} \overline{\ps} (\Ga)^{k+1} (i \prt)^k \ch +  
h.c.
\quad ,
\label{cptbroken}
\eeq
where $\la$ is a dimensionless coupling constant, $\ch$ and $\ps$  
denote generic fermionic fields, $\Ga$ denotes a gamma-matrix  
structure, $(i\prt)^k$ represents the action of a four-derivative at  
order $k \ge 0$ and the Lorentz indices were suppressed for  
simplicity. As before, Lorentz and now CPT symmetries are violated as
$T$ acquires non-vanishing vacuum expectation values $\vev{T}$. From  
\rf{cptbroken} one sees that factors of $i\prt_0$ also introduce a  
suppression at low-energies.

An estimate of the violation of the Lorentz symmetry can be obtained  
from the
effect of the interactions \rf{cptbroken} between tensor field  
fluctuations around (3), generically denoted by $h$, to for instance a  
nucleon field, $N$, with momentum $p$:
\beq
\lambda <T> \left({p \over M} \right)^{k}
{h \over M} \bar N N
\quad ,
\label{lorentzsize}
\eeq
where we have scaled $h$ by $<T>$ and assumed that the appropriate  
contraction of indices is implied. Since in most experiments $p <<  
M$, hence for $m_l \sim p$, the condition $k+ l \geq 1$ has to be  
respectd in order to avoid obvious contradictions with existing  
experimental tests (see below). We can now study the set of values  
consistent with the measured deviations from the Lorentz symmetry.  
Notice, however that these considerations imply that a direct  
detection of the interaction (6) is experimentally ruled out as its  
coupling strength $\alpha =\lambda (m_1 /M)^{k+1}$ is well within the  
region that, due to Newtonian and electrostatic background effects,  
is at present inaccessible for $\alpha \leq 10^{-2}$ when $\lambda  
\sim 10^{-4}m$ \cite{beane}.

Limits on the violations of the Lorentz symmetry have been searched  
via
laser-interferometric versions of the Michelson-Morley experiment  
that allow a comparison between the velocity of light, $c$, and the  
maximum attainable velocity of massive particles, $c_0$, up to  
$\delta \equiv |c^2/c_{0}^2 - 1| < 10^{-9}$ \cite{bhall}, and  
through the measurement
of the quadrupole splitting time dependence of nuclear Zeeman levels
along Earth's orbit. Experiments of the latter nature, known as  
Hughes-Drever tests \cite{hrbl, drever}, have been performed over the  
years \cite{prestage, lamoreaux}, the most recent one \cite{chupp} revealing that  
$\delta < 3 \times 10^{-21}$~\footnote{In these tests, deviations from  
the Local Lorentz Invariance are inferred via possible anisotropies  
of the inertial mass, $m_I$, and $\delta = |m_I c^2 /\sum_A E^A -  
1|$, where the sum is over all forms of internal energy of a chosen  
nucleus.}. Limits on the violation of the momentum conservation and  
existence of
a preferred reference frame can be also inferred from the limits on  
the
parametrized post-Newtonian parameter $\alpha_{3}$ obtained
from the pulse period of pulsars \cite{will} and millisecond  
pulsars \cite{bell}.
This parameter vanishes identically in general relativity and the  
recent bound $|\alpha_{3}| < 2.2 \times 10^{-20}$ obtained from  
binary pulsar systems \cite{bd} implies the Lorentz symmetry holds 
up to the level established by that limit.

Let us now estimate the impact at low-energy of the string  
interactions.
For the most stringent experimental tests of the Lorentz invariance,
the Hughes-Drever tests, $m_l \sim MeV (>> p)$ and we see from
\rf{lorentzsize} that Lorentz invariance violating effects are
consistent with the experimental limits for $k=0$ and $l=1$. We  
mention that these conclusions are also consistent (for $k=1$ in this 
instance) with the recently  
discussed bounds for $\delta$ that can be obtained from highly  
energetic cosmic rays ($E \sim 10^{20}~eV$) and from neutrino  
oscillations \cite{cg}. We also point out that as far as violation  
of the CPT symmetry is concerned, the relevant condition when  
confronting with experimental evidence is $k+l > 1$ \cite{kp2}.  
Thus, if we aim to account for both Lorentz and CPT violations, we  
should take $k=0$ and $l=2$. This would imply that Lorentz symmetry  
breaking effects are well below the existing experimental limits,  
although they can still saturate the vacuum energy in this case as  
well.

Thus, we have seen that non-trivial solutions in string filed theory  
may lead
to fairly new implications and, in particular, to the quite distinct  
phenomenological signatures such as the ones associated with the  
violations of  Lorentz and CPT symmetries. The spontaneous breaking  
of Lorentz symmetry arising from string field theory does imply that  
Lorentz tensor fields acquire masses and from that follows the  
existence of new tensor Yukawa type interactions. Assuming that the  
contribution of the vacuum expectation values of Lorentz
tensors to the vacuum energy are uncancelled and responsible for the
observed bounds to the cosmological constant implies that the range  
of the new interaction is $\lambda \sim  10^{-4} \ m$.

We would like to close with two final remarks. The discussed experimental  
limits on preferred-frame effects affect naturally the cosmological  
constant itself as the absence of the exact Lorentz symmetry implies  
its value is not constant throughout space. This feature leads to an  
anti-bias factor: structure formation occurs preferably where  
$\Lambda$ is smaller. Finally, in a quite recent paper, Martel,  
Shapiro and Weinberg \cite{msw} obtained, assuming the   
cosmological constant takes different values at different sites  
(referred to as subuniverses), that $\rho_V \leq 3\rho_{Matter}$ is under
fairly reasonable assumptions the most  
likely value for the vacuum energy. This value for  
$\rho_V$ as well as the line of reasoning of that article could, of  
course, very well be used in the present work.


\newpage


\begin{thebibliography}{99}


\baselineskip = 18pt

\bibitem{manton} N.S. Manton, Nucl. Phys. {\bf B158} (1979) 141.


\bibitem{dp} M.J. Duff and C.N. Pope, Nucl. Phys. {\bf B255} (1985)  
355.


\bibitem{kmrv} Yu.A. Kubyshin, J.M. Mour\~ao, G. Rudolph and I.P.  
Volobujev,
Lecture Notes in Physics {\bf 349} (Springer Verlag, 1988)


\bibitem{kz} D. Kapetanakis and G. Zoupanos, Phys. Rep. {\bf C219}  
(1992) 1.


\bibitem{ks1} V.A. Kosteleck\'y and S. Samuel,
Phys. Rev. {\bf D39} (1989) 683.


\bibitem{ks2} V.A. Kosteleck\'y and S. Samuel,
Phys. Rev. Lett. {\bf 63} (1989) 224.


\bibitem{ds} M. Dine and N. Seiberg, Phys. Lett. {\bf B162} (1986)  
299.


\bibitem{filq} A. Font, L.E. Ib\'a\~nez, D. L\"ust and F. Quevedo,  
Phys. Lett. {\bf B245} (1990) 401; {\bf B249} (1990) 35.

\bibitem{phillips} P.R. Phillips, Phys. Rev. {\bf 146} (1966) 966.

\bibitem{nn} H.B. Nielsen and M. Ninomiya, Nucl. Phys. {\bf B141}  
(1978) 153.


\bibitem{weinberg1} S. Weinberg, Phys. Lett.
{\bf B138} (1984) 47.


\bibitem{witten1} E. Witten, Nucl. Phys. {\bf B186} (1981) 412;
{\bf B195} (1982) 481.


\bibitem{kp1}
V.A. Kosteleck\'y and R. Potting,
Nucl. Phys. {\bf B359} (1991) 545;
Phys. Lett. {\bf B381} (1996) 389.

\bibitem{kp2}
V.A. Kosteleck\'y and R. Potting,
Phys. Rev. {\bf D51} (1995) 3923.


\bibitem{ckv}
D. Colladay and V.A. Kosteleck\'y,
Phys. Lett. {\bf B344} (1995) 259;
Phys. Rev. {\bf D52} (1995) 6224;
V.A. Kosteleck\'y and R. Van Kooten,
Phys. Rev. {\bf D54} (1996) 5585.

\bibitem{adler} R. Adler et al. (CPLEAR Collaboration), J. Ellis, J.L. Lopez,
N.E. Mavromatos and D.V. Nanopoulos,
Phys. Lett. {\bf B364} (1995) 239.


\bibitem{bckp} O. Bertolami, D. Colladay, V.A. Kosteleck\'y
and R. Potting, Phys. Lett. {\bf B395} (1997) 178.


\bibitem{witten2} E. Witten, Nucl. Phys. {\bf B268} (1986) 253.


\bibitem{weinberg2} S. Weinberg, Rev. Mod. Phys.
{\bf 61} (1989) 1.


\bibitem{weinberg3} S. Weinberg, ``Theories of the Cosmological Constant''
(astro-ph9610044).


\bibitem{coleman} S. Coleman, Nucl. Phys. {\bf B307} (1988) 867.


\bibitem{bb} M.C. Bento and O. Bertolami, Gen. Rel. and Gravitation
{\bf 28} (1996) 565.


\bibitem{witten3} E. Witten, Int. J. Mod. Phys. {\bf 10} (1995) 1247.


\bibitem{moore} G. Moore, Nucl. Phys. {\bf B293} (1987) 139.


\bibitem{cfkrr} G.D. Coughlan, W. Fischler, E.W. Kolb, S. Raby and
  G.G. Ross, Phys. Lett. {\bf B131} (1983) 59.


\bibitem{enq} J. Ellis, D.V. Nanopoulos and M. Quir\'os,
Phys. Lett. {\bf B174} (1986) 176.


\bibitem{bertolami} O. Bertolami and G.G. Ross, Phys. Lett. {\bf B183} (1987) 
163.


\bibitem{bertolami} O. Bertolami, Phys. Lett. {\bf B209} (1988) 277.


\bibitem{ccqr} B. de Carlos, J.A. Casas, F. Quevedo and E. Roulet,
Phys. Lett. {\bf B318} (1993) 447.


\bibitem{bb1} M.C. Bento and O. Bertolami, Phys. Lett. {\bf B336}  
(1994) 6.


\bibitem{bkn} T. Banks, D.B. Kaplan and A.E. Nelson,
 Phys. Rev. {\bf D49} (1994) 779.


\bibitem{bbs} T. Banks, M. Berkooz and P.J. Steinhardt,
Phys. Rev. {\bf D52} (1995) 705.


\bibitem{bbsms} T. Banks, M. Berkooz, S.H. Shenker, G. Moore and P.J.
Steinhardt, Phys. Rev. {\bf D52} (1995) 3548.


\bibitem{esm} G. Efstathiou, W.J. Sutherland and S.J. Maddox, Nature
{\bf 348} (1990) 705.


\bibitem{bh} M. Bolte and C. Hogan, Nature
{\bf 376} (1995) 399.


\bibitem{rpk} A.G. Riess, W.H. Press and R.P. Kirshner, Ap. J.
{\bf 438} (1995) L17.


\bibitem{freedman} W.L. Freedman at al., Nature
{\bf 371} (1994) 757.


\bibitem{ft} M. Fukugita and E.L. Turner, MNRAS
{\bf 253} (1991) 99.


\bibitem{mr} D. Maoz and H-W. Rix, Ap. J.
{\bf 416} (1993) 425.


\bibitem{os} J.P. Ostriker and P.J. Steinhardt, Nature
{\bf 377} (1995) 600.


\bibitem{perlmutter} S. Perlmutter et al., ``Measurement of the
Cosmological Parameter $\Omega$ and $\Lambda$ from the First
7 Supernovae at $z \ge 0.35$'' (astro-ph/9608192).


\bibitem{ct} R.G. Crittenden and N. Turok, ``Looking for $\Lambda$
with the Rees-Sciama effect'' (astro-ph/9510072).


\bibitem{kt} L.M. Krauss and M.S. Turner, Gen. Rel. and Gravitation
{\bf 27} (1995) 1137.


\bibitem{fit} E.Fishbach and C. Talmadge, ``Ten Years of the Fifth  
Force'' (hep-ph/9606249).


\bibitem{beane} S.R. Beane, ``On the importane of testing gravity at
distances less than $1 \ cm$''
(hep-ph/9702419).


\bibitem{price} J.C. Price, Proceedings of the International  
Symposium
on Experimental Gravitational Physics, eds. P. Michelson, H. En-Ke  
and
G. Pizzella (D. Reidel, Dordrecht 1987).

\bibitem{dg} S. Dimopoulos and G.F. Guidice, Phys. Lett.
{\bf B379} (1996) 105.


\bibitem{bhall} A. Brillet and J.L. Hall, Phys. Rev. Lett.
{\bf 42} (1979) 549.


\bibitem{hrbl} V.W. Hughes, H.G. Robinson and V. Beltran-Lopez,
Phys. Rev. Lett. {\bf 4} (1960) 342.


\bibitem{drever} R.W.P. Drever, Philos. Mag.
{\bf 6} (1961) 683.


\bibitem{prestage} J.D. Prestage, J.J. Bollinger, W.M. Itano and
D.J. Wineland, Phys. Rev. Lett.
{\bf 54} (1985) 2387.


\bibitem{lamoreaux} S.K. Lamoreaux, J.P. Jacobs, B.R. Heckel, F.J.  
Raab and
E.N. Fortson, Phys. Rev. Lett.
{\bf 57} (1986) 3125.


\bibitem{chupp} T.E. Chupp, R.J. Hoare, R.A. Loveman, E.R. Oteiza,
J.M. Richardson M.E. Wagshul and A.K. Thompson, Phys. Rev. Lett.
{\bf 63} (1989) 1541.


\bibitem{will} C.M. Will, ``Theory and Experiment in Gravitational  
Physics''
(Cambridge University Press, 1993).


\bibitem{bell} J.F. Bell, Ap. J. {\bf 462} (1996) 287.


\bibitem{bd} J.F. Bell and T. Damour, Class. Quantum Gravity
{\bf 13} (1996) 3121.


\bibitem{cg} S. Coleman and S.L. Glashow, ``Cosmic Ray and Neutrino  
Tests of Special Relativity'' (hep-ph/9703240).


\bibitem{msw} H. Martel, P.R. Shapiro and S. Weinberg, ``Likely  
Values of the Cosmological Constant'' (astro-ph/9701099).

\end{thebibliography}
\end{document}